\documentclass[aps,pra,twocolumn,longbibliography,superscriptaddress,floatfix]{revtex4-1}
\usepackage{amsfonts}
\usepackage[dvipdfmx]{graphicx}
\usepackage{epsfig}
\usepackage{dcolumn}
\usepackage{bm}
\usepackage{amsmath}
\usepackage{graphicx}
\usepackage[latin1]{inputenc}
\usepackage{ulem}
\usepackage{epstopdf}
\usepackage{subfigure}
\usepackage{color}
\usepackage{amsthm}
\usepackage{newlfont}
\usepackage{graphicx}
\usepackage{epstopdf}
\usepackage{appendix}
\usepackage[breaklinks=true]{hyperref}
\usepackage{breakcites}
\usepackage{textcomp}
\usepackage{appendix}
\usepackage{multirow}
\usepackage{color}
\usepackage{amssymb}
\usepackage{epsfig}
\usepackage{mathptmx}
\usepackage{bm}
\usepackage[american]{babel}
\usepackage{braket}

\hypersetup{colorlinks=true,linkcolor=blue,citecolor=blue,filecolor=blue,
urlcolor=blue,pdfstartview=FitH}
\newcommand{\be}{\begin{equation}}
\newcommand{\ee}{\end{equation}}
\newcommand{\bs}{\begin{split}}
    \newcommand{\es}{\end{split}}
\newcommand{\bea}{\begin{eqnarray}}
\newcommand{\eea}{\end{eqnarray}}

\begin{document}

\title{Topological and flat band states induced by hybridized linear interactions in one-dimensional photonic lattices}

\author{G. C\'aceres-Aravena}
\affiliation{Departamento de F\'isica and MIRO, Facultad de Ciencias
F\'isicas y Matem\'aticas, Universidad de Chile, Santiago, Chile}
\author{L. E. F. Foa Torres}
\affiliation{Departamento de F\'{\i}sica, Facultad de Ciencias F\'isicas y Matem\'aticas, Universidad de Chile, Santiago, Chile}
\author{R.A. Vicencio}
\affiliation{Departamento de F\'isica and MIRO, Facultad de Ciencias
F\'isicas y Matem\'aticas, Universidad de Chile, Santiago, Chile}

\date{\today}

\begin{abstract}

We report on a study of a one-dimensional linear photonic lattice hosting, simultaneously, fundamental and dipolar modes at every site. We show how, thanks to the coupling between different orbital modes, this minimal model exhibits rich transport and topological properties. By varying the detuning coefficient we find a regime where bands become flatter (with reduced transport) and, a second regime, where both bands connect on at a gap-closing transition (with enhanced transport). We detect an asymmetric transport due to the asymmetric inter-mode coupling and a linear energy exchange mechanism between modes. Further analysis show that the bands have a topological transition with a non-trivial Zak phase which leads to the appeareance of edge states in a finite system. Finally, for zero detuning, we found a symmetric condition for coupling constants, where the linear spectrum becomes completely flat, with states fully localized in space occupying only two lattice sites.

\end{abstract}

\maketitle

\section{Introduction}

During the last twenty five years, the study of light propagation in periodic media, which mimics solid-state physics systems, has successfully shown several fundamental phenomena with potential applications~\cite{rep1,rep2,rep3}. This research area has not been limited to waveguide arrays, expanding also to cold atoms~\cite{bec0}, ferromagnetic lattices~\cite{solidstate}, microcavity coupled systems~\cite{amo1}, phononic lattices~\cite{aco1}, among others. Photonic lattices, in particular, have concentrated enormous attention due to its flexibility to fabricate one (1D) and two (2D) dimensional coupled systems~\cite{guiasalex}, where several predictions from condensed matter, including Anderson Localization~\cite{segev1}, Topological Insulation~\cite{rech1} and Flat Band (FB) phenomena~\cite{fbprl,fbprlseba}, have found a privileged testing ground.

Most of the studies in photonic lattices consider the simplest excitation of a lattice; i.e., the fundamental ``s'' wavefunction dynamics. However, recent advances in photonics allow the excitation of higher orbital states either by using spatial light modulator devices~\cite{dipole1,graphene1} or by selecting the associated energy of a given spectrum~\cite{amo19}. Thanks to this progress one can now also investigate the interaction between ``s'' and ``p'' states~\cite{mangussi19}, where a wealth of new phenomena have been predicted~\cite{mag1,bec2,bec3,bec4,slot}. For example, the isolated excitation of ``p'' states on a graphene lattice could generate the appearance of flat-band phenomena due to a perfect cancellation of transport when correctly choosing the dipole orientation on a given ring~\cite{bec1,amo2}. By assuming hybridized linear interactions, we recently found that a twisted 1D lattice presents flat-band phenomena as well as edge, compact and exponential, states~\cite{gab19}. It is thus interesting to explore the interplay between the lattice configuration and the interaction between different orbitals. Depending on their parity and symmetry, the  degree of freedom due to additional orbitals could induce non-trivial coupling and, therefore, non-trivial band phenomenology with non-standard transport and localization properties.

Another research front merging and synergizing with the mentioned studies is that of topological insulators~\cite{asboth16,ortmann15,hasan10}. Topological photonics~\cite{khanikaev13,ozawa19} is an emerging field seeking to exploit photonic lattices not only to realize and probe topological states, but also to realize truly new effects unknown in their parent communities (i.e., condensed matter and photonics). In this context one may also seek to generate non-trivial topological states from the coupling between higher-order states, as we will show later in this paper. 

From a more applied point of view, it has been shown that FB lattices are special candidates for using waveguide arrays as key elements for optical communications~\cite{kagome1,stub,chen1,chen2,OLsaw,OLdiamond,patent}, while the use of orthogonal states could help to improve spatial multiplexing techniques~\cite{multi1}. For example, FB localization occurs at linear level and without requiring any change in the lattice geometry. This allows the combination between different compact states which forms a completely coherent pattern, which propagates stable along the propagation coordinate and for any level of power. In general, FB lattices are formed by systems possessing at least two bands~\cite{rep3,FBluis}, which have a special geometric configuration that makes possible the occurrence of destructive interference effects. This amplitude annihilation exactly cancels the transport across the lattice, generating FB linear states which are perfectly compact with a zero tail.

In this paper, we present the simplest possible 1D lattice model for which the coupling between ``s'' and ``p'' states allow the excitation of interesting properties in terms of transport, localization, energy exchange, topology, and flat-band phenomena. Our findings include: i) a tunable transition from quasi-flat bands (with reduced transport) to one with more dispersive bands with enhanced transport; ii)  the observation of asymmetric transport produced by asymmetric inter-mode coupling and a linear energy exchange mechanism between modes; iii) a topological transition appearing at a gap closing in parameter space supported by calculation of the Zak phase and visualization of the associated edge states; iv) the emergence of a completely flat band system with compact localized states under specific conditions. The lattice proposed in this work could be easily fabricated by current experimental photonic techniques~\cite{amo1,guiasalex}, making possible a direct observation of our theoretical predictions.
%
\begin{figure*}
\centering
\includegraphics[width=1.7\columnwidth]{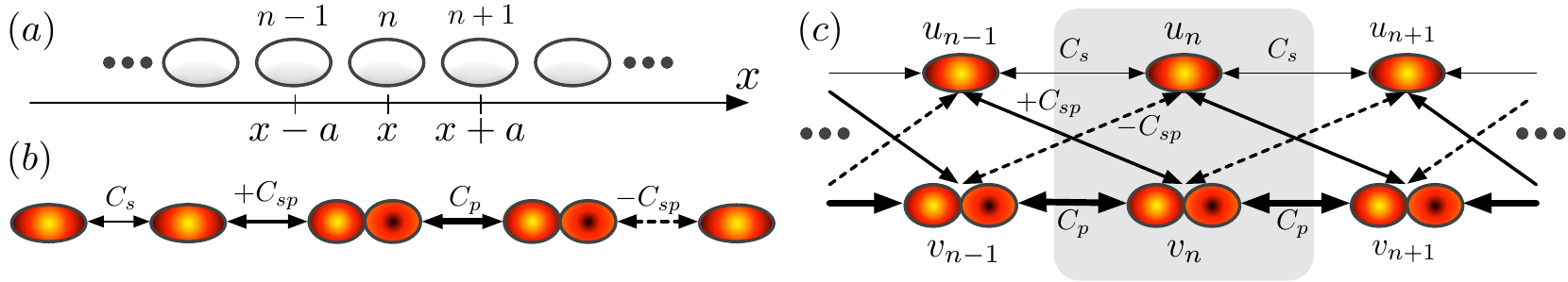}
\caption{(a) A 1D lattice of horizontally oriented elliptical waveguides. (b) On-row coupling interactions. (c) Effective model, with an unitary cell shown in grey.}\label{f1}
\end{figure*}
%

\section{The model}

We explore one of the simplest one-dimensional configuration to study hybridized coupling interactions in photonic lattices. Our model consists on a row of identical waveguides which are disposed horizontally, as shown in Fig.~\ref{f1}(a). This orientation allows us to study a lattice presenting coupling between fundamental and dipolar modes, which is absolutely absent in standard vertically oriented configurations~\cite{dipole1}. Typical fabrication methods in optics, based on an axial focusing of a femtosecond laser on a glass substrate~\cite{guiasalex}, produce elliptically oriented waveguides. In that context, we have experimentally proved that the excitation of dipolar states is possible only in the elongated waveguide direction, for visible wavelengths~\cite{dipole1,graphene1}. Therefore, in order to effectively study a coupling mechanism between these type of modes, we select a simple 1D configuration.

If we assume that ``s'' and ``p'' modes can be excited simultaneously at every waveguide of the lattice~\cite{com1}, we could draw a coupling-interaction picture, considering nearest-neighbor sites and different modes, as the one sketched in Fig.~\ref{f1}(b). We directly consider linear interactions between modes of different waveguides, having in mind a standard superposition integral~\cite{poladiso}
$$
C_{nm}(a)\sim \int_{-\infty}^{+\infty} \phi_m (x+a,y)\phi^*_n(x,y)\ dxdy\ ,
$$
for a distance ``$a$'' between waveguides. $\phi_n$ and $\phi_m$ describe the spatial profile of a given mode at sites $n$ and $m$, respectively. 
The integral is performed over the transversal area and corresponds to the coupling constant between different modes at different waveguides (the coupling between $s$ and $p$ modes at the same waveguide is exactly zero due to the orthogonality of wavefunctions). The usual nearest-neighbor coupling between equal modes is defined as $C_s$ and $C_p$, for $s$ and $p$ modes~\cite{gab19}, respectively. In our model, the superposition between fundamental modes is such that $C_s>0$, while the coupling between dipolar states is always negative ($C_p<0$). The sign of coupling between fundamental and dipolar states depends on the particular position of wave-functions. As we show in Fig.~\ref{f1}(b), a fundamental mode couples to a dipolar one to the right via a positive coefficient $+C_{sp}$, while this coupling becomes negative ($-C_{sp}$) when the direction is reversed. In general, due to the spatial extension of modes~\cite{dipole1}, $|C_s|<|C_{sp}|<|C_{p}|$ for a given distance ``$a$'', which corresponds to the lattice constant. 

In our system, the light evolution occurs along the $z$ direction (the dynamical variable in our model), and the lattice dynamics is well described by a set of coupled discrete linear Schr\"{o}dinger equations~\cite{rep1,rep2} that, considering the effective lattice system sketched in Fig.~\ref{f1}(c), read as

\begin{eqnarray}
-i\frac{d u_n}{d z} = \beta_{s} u_{n}+C_s (u_{n+1}+u_{n-1})+C_{sp}(v_{n+1}-v_{n-1})\ ,\nonumber\\
-i\frac{d v_n}{d z} = \beta_{p} v_{n}+C_p (v_{n+1}+v_{n-1})-C_{sp}(u_{n+1}-u_{n-1})\ .\nonumber \\ 
\label{mo2}
\end{eqnarray}
Here, $u_n$ and $v_n$ correspond to the light amplitude of $s$ and $p$ modes, respectively, at the $n$-th site of a full 1D lattice. $\beta_i$ corresponds to the respective on-site propagation constants, while coupling interactions are well described in Fig.~\ref{f1}(c).

Although this model bears similarities with Creutz's model~\cite{Creutz}, which has been experimentally realized recently in ultracold fermions~\cite{Kang}, in our case we do not have direct links between the two modes and time-reversal symmetry is preserved. Furthermore, we note that our model is non-interacting, unlike the Creutz-Hubbard model considered in Ref.~\cite{June}.

\section{Band analysis and transport}

Stationary solutions are of the form $\{u_n(z),v_n(z)\}=\{u_n,v_n\}\exp [i (\lambda+\beta_p) z]$, where we have added a phase transformation to simplify equations; after inserting this ansatz into model (\ref{mo2}), we obtain the following set of coupled equations
\begin{eqnarray}
\lambda u_n = \Delta \beta u_n+C_s (u_{n+1}+u_{n-1})+C_{sp}(v_{n+1}-v_{n-1})\ ,\nonumber\\
\lambda v_n = C_p (v_{n+1}+v_{n-1})-C_{sp}(u_{n+1}-u_{n-1})\ .\hspace{1.37cm}
\label{mo3}
\end{eqnarray}
Here, $\lambda$ corresponds to the longitudinal propagation constant of the coupled system, which defines the propagation properties along the $z$-direction for global solutions of this model ($\lambda$ is also known as the solution's frequency in $k$-space). $\Delta \beta\equiv\beta_s-\beta_p$ defines the detuning between $s$ and $p$ modes, which is typically larger than zero for optical waveguides.

Now, we look for the linear spectrum of model (\ref{mo3}) by using a plane wave (Bloch-like) ansatz $\{u_n,v_n\}=\{U,V\}\exp (i k_x n a)$, with $k_x$ the transversal wave-vector. By doing this, we find the following spectrum for propagating linear waves,

\begin{equation}
\lambda_{\pm}(k_x) =\Delta \beta/2 +(C_s+C_p)\cos (k_x a)\pm \sqrt{g(k_x)}\ ,
\label{dr1}
\end{equation}

\begin{figure}[t]
\begin{center}
\includegraphics[width=0.48\textwidth]{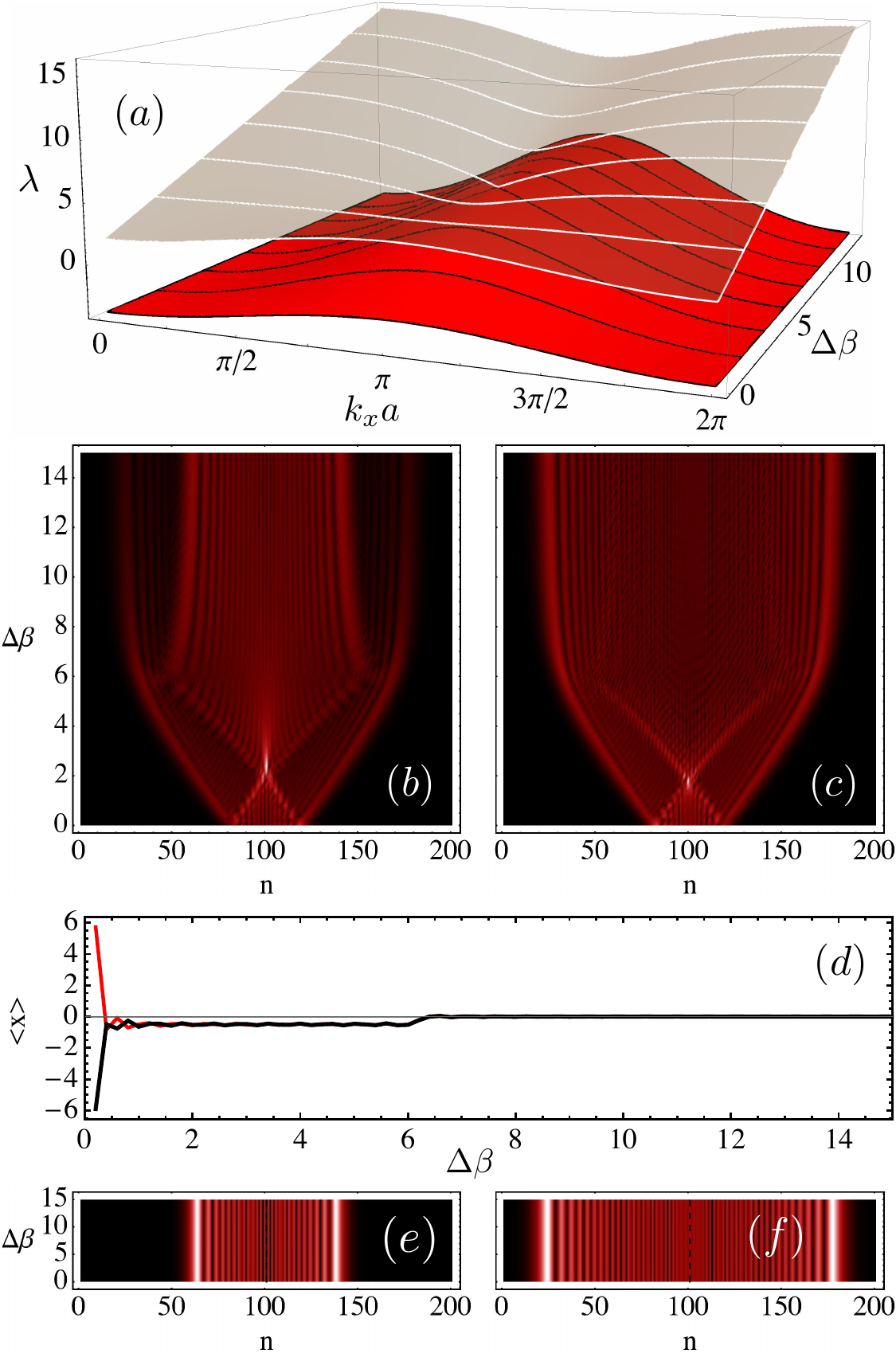}
\caption{(a) $\lambda_{\pm}$ versus $k_x a$ and $\Delta \beta$, where guidelines are located at $\Delta\beta=0,2,4,6,8,10,12,14$. (b) and (c) Output intensity profiles at $z_{max}=20$ versus $\Delta \beta$ for ``s'' and ``p'' modes, respectively. (d) Output center of mass for profiles shown in (b) and (c), using red and black lines, respectively. $\{C_s,C_p,C_{sp}\}=\{1,-2,1.5\}$. (e) and (f) Output intensity profiles at $z_{max}=20$ versus $\Delta \beta$ for ``s'' and ``p'' modes, respectively, for $\{C_s,C_p,C_{sp}\}=\{1,-2,0\}$.}\label{f2}
\end{center}
\end{figure}

where $$g(k_x) =[\Delta \beta/2+(C_s-C_p) \cos (k_x a)]^2+4 C_{sp}^2 \sin^2 (k_x a)\ .$$ These two bands are presented in Fig.~\ref{f2}(a), as a function of $k_x a$ and $\Delta\beta$, for a given set of standard parameters. First of all, we notice that the lower band always has a minimum value of $\lambda_-=2C_p$, at the border of the Brillouin zone. Then, we observe that for a small detuning $\Delta \beta$, the two bands are closer and also flatter, indicating a reduced transport across the lattice. 
We numerically find that for $\Delta \beta\approx 2$ the upper band width decreases to a minimum, but it is not completely flat due to the non trivial curvature this band shows at that region. Below $\Delta \beta\sim 2$ the upper band has a simple maximum in the interval $k_x a\in\{0,2\pi\}$, while this panorama changes above $\sim 2$ where two maxima and one minimum appear (the expression for this transition is not analytically trivial, so we simply do not write it here). Although the upper band does not become perfectly flat, a reduced transport must be observed at that region of parameters. By further increasing the detuning coefficient, we observe that there is a special region where the two bands touch and form a non-symmetric Dirac-like cone. It is straightforward to obtain the value for $\Delta \beta$ in which this happens. We simply ask for a set of parameters where $\lambda_+=\lambda_-$, obtaining the solution $k_x=\pi/a$ and $\Delta \beta=2(C_s-Cp)$, where both bands have a value $-2C_p$ (for parameters used in Fig.~\ref{f2}, $\Delta \beta=6$ and $\lambda_+=\lambda_-=4$). In this region, we expect to observe an increasing speed for travelling waves, due to the slope divergency at this special point. This transition point might also host a topological transition~\cite{asboth16,ortmann15,hasan10} for the associated bands. In the next section we clarify this issue. 

We notice that for $\Delta\beta\geqslant 2(C_s-Cp)$, the lower band saturates having a maximum value of $\lambda_-=-2C_p$, at $k_x=\pi/a$. Therefore, the transport due to the lower band saturates too, without any important effect for an increasing detuning coefficient. The lower band gets bounded to the region $\lambda_-\in\{2C_p,-2C_p\}$, acting as a one-dimensional isolated system for ``p'' modes, with $\lambda_-\approx 2C_p\cos(k_x a)$. The upper band increases almost linearly for larger $\Delta\beta$, observing also the formation of an effective one-dimensional system for ``s'' modes, with $\lambda_+\approx \Delta\beta+ 2C_s\cos(k_x a)$. As a consequence, we expect a saturation of transport, in terms of expansion, for both modes due to the effective dynamical decoupling between both states.

\subsection{Single-site excitation}

Now, we study transport phenomena by numerically integrating model (\ref{mo2}), considering different input conditions, a propagation distance $z_{max}=20$ and a lattice of $N=201$ sites. First of all, we excite our lattice using a bulk single-site excitation at $n_i=101$, simultaneously for ``s'' and ``p'' modes. Figs.~\ref{f2}(b) and (c) show the output profile versus $\Delta \beta$ for both states, respectively. A single-site excitation excites several linear modes in the lattice, which propagate depending on their respective $k_x$-vectors, producing a typical discrete diffraction pattern (characterized by two external main lobes away from the input site region~\cite{rep1}). A single-site excitation will excite modes from both bands, but selectively. This means that the excitation of fundamental (dipolar) modes will excite the upper (lower) bands more efficiently. This is observed in Figs.~\ref{f2}(b) and (c), where a smaller diffraction area is obtained for ``s'' compared to ``p'' modes, respectively. We can also observe the expression of linear bands on the dynamics. We clearly observe a tendency to localize energy for $\Delta \beta\approx 2$, where both modes show a bright spot plus a background radiation. This effect is obviously stronger for the ``s'' mode, which is the one related to the upper band that becomes flatter close to this detuning value. This absence of transport plus radiation is quite similar to the experimental evidence found for Sawtooth lattices~\cite{OLsaw}, where close to the FB condition there is a flatter and a dispersive band excited simultaneously.

For $\Delta \beta >2$, we observe a tendency to increase the diffraction area, for both modes, due to an increment of the band dispersion. We observe  in Figs.~\ref{f2}(b) [(c)] that the ``s'' (``p'') mode diffracts with a smaller (larger) maximum velocity due to a stronger excitation of the upper (lower) band. The enhancement in the transport saturates around the Dirac cone region. As explained above, this is because the band dispersion does not increase any further, observing some kind of 1D effective isolated transport, with corresponding coupling coefficients, and with some small interaction between modes due to a non-zero $C_{sp}$ coefficient. In Fig.~\ref{f2}(d) we show the output center of mass, defined as
\begin{equation*}
<x>\ =\frac{\sum_n n |w_n(z_{max})|^2}{\sum_n |w_n(z_{max})|^2}-n_i\ , 
\end{equation*}
for $w_n=u_n,v_n$ and $n_i$ the input site. We observe a non-zero evolution of this quantity, which is highly connected to the corresponding dispersion relation and the sine-function dependence of it. Standard lattices are mostly governed by cosine-like functions, which are completely symmetric around $k_x=0$. Here, the coupling $C_{sp}$ in model (\ref{mo2}) is not symmetric and some asymmetries in the transport are expected, also as a direct consequence of having positive and negative coupling interactions. We observe that for small $\Delta \beta$ the ``s'' mode shifts the center of mass to the right, while the ``p'' mode does to the left quite symmetrically. Then, for $\Delta \beta\gtrsim 0.5$ both modes are shifted to the left up to the Dirac cone region, where both symmetrize with respect to the input position $n_i$. This symmetrization is also a consequence of the effective isolated 1D dynamics for both modes for an increasing value of the detuning. However, the $<x>$ values are close to zero but not exactly zero, indicating the impact of the nonzero $C_{sp}$ coefficient [a null $C_{sp}$ value simply decouples the equations and both dynamics are completely symmetric with respect to the input position, as shown in Figs.~\ref{f2}(e) and (f)].

\subsection{Gaussian excitation}
%
\begin{figure}[t]
\begin{center}
\includegraphics[width=0.48\textwidth]{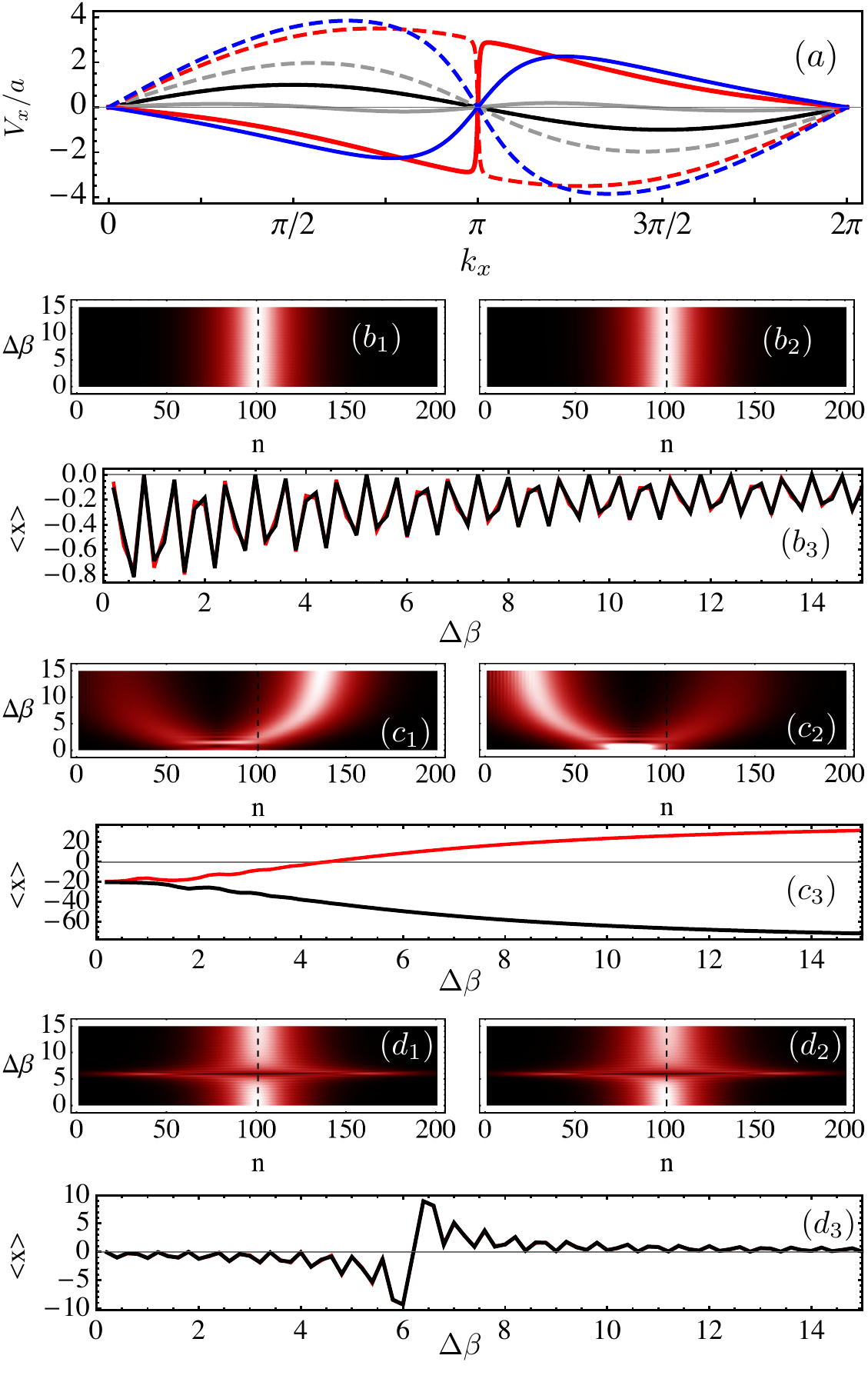}
\caption{(a) $V_x/a$ versus $k_x$ for $\Delta\beta=0$ (black), $2$ (gray), $6$ (red), and $10$ (blue), for the upper (full lines) and lower (dashed lines) bands. Output intensity profiles at $z_{max}=20$ versus $\Delta \beta$, for (s,p) modes: (b1,b2) for $k_x=0$, (c1,c2) for $k_x=\pi/2$, and (d1,d2) for $k_x=\pi$, respectively. (b3), (c3), and (d3) Output center of mass for profiles shown in (b), (c) and (d), using red and black lines for ``s'' and ``p'' modes, respectively. $\{C_s,C_p,C_{sp}\}=\{1,-2,1.5\},\ \alpha=0.002$ and $n_i=101$.}\label{f3}
\end{center}
\end{figure}

Before exploring gaussian beam propagation in our lattice, we compute the transversal velocity defined as: $V_x/a=\partial\lambda/\partial k_x$ for both bands. We plot this quantity in Fig.~\ref{f3}(a) for four $\Delta\beta$ values. We observe how the slope of each band defines the velocity of plane waves. For $\Delta\beta=0$ both band behaves equal (black curve), while for $\Delta\beta=2$ the upper band shows a very small velocity (quasi-flat band) with a lower band having a maximum spreading at $k_x=\pi/2$. Then, for an increasing $\Delta\beta$-value we observe a clear separation between both bands, having a positive (negative) velocity the lower (upper) band waves in the interval $k_x\in\{0,\pi\}$. For $\Delta\beta=6$, we clearly observe the Dirac cone phenomenology with a diverging velocity at $k_x=\pi$. In general, we observe that the transport in this lattice depends on the specific excited band; however, the inter-mode coupling ($C_{sp}$) also generates a more complex dynamics due to a power exchange mechanism between both modes during propagation. This kind of interaction is common for vector-like systems~\cite{ve1,ve2}, where a nonlinear four-wave mixing mechanism is responsible for this. Although model (\ref{mo2}) is linear, we can write an expression for a power exchange between modes $$\frac{\partial P_u}{\partial z}=-\frac{\partial P_v}{\partial z}=-2C_{sp}\sum_n \left[Im (v_{n+1}u_n^*)+Im (u_{n+1}v_n^*)\right]\ ,$$ with $P_u\equiv\sum_n |u_n|^2$ and $P_v\equiv\sum_n |v_n|^2$, which is originated due to the asymmetric coupling between ``s'' and ``p'' modes. As $C_{sp}>0$, the fundamental mode always releases some energy to the dipolar state at the beginning. This originates an oscillation of energy between modes, depending on the accumulated phase during the dynamics. Therefore, the dynamics becomes complex due to the excited bands and due to this energy exchange mechanism.

We explore plane-wave propagation by numerically initializing model (\ref{mo2}) by a Gaussian input excitation of the form: $\{u_n(0),v_n(0)\}=\exp [-\alpha(n-n_i)^2]\exp [i k_x (n-n_i) a]$. Here, $\alpha$ is the gaussian exponent and $k_x$ the transversal momentum. 
In Fig.~\ref{f3} we show some examples. For $k_x=0$ we observe in Figs.~\ref{f3}(b1) and (b2) almost no transversal transport for both modes, as expected from the null velocities at this $k_x$-value [see Fig.~\ref{f3}(a)]. However, the output center of mass [Fig.~\ref{f3}(b3)] shows a slight shift from the input position. This is due to the nonzero $C_{sp}$ coupling that causes some asymmetries in the interaction between both states, due to the power exchange mechanism. Then, we increase the momentum to $k_x=\pi/2$ as Figs.~\ref{f3}(c1) and (c2) show. We observe that for small $\Delta\beta$-values the ``p'' state is larger in intensity (however, it is important to mention that this is just a fixed output picture at $z_{max}=20$) and both states are shifted to the left as Fig.~\ref{f3}(c3) indicates. Then, close to $\Delta\beta=2$ the ``s'' mode has a larger intensity due to the flat-band tendency. For larger $\Delta\beta$-values the ``s'' mode tends to move strongly to the right while the ``p'' mode keeps moving strongly to the left. However, both states have also a smaller intensity propagating beam (a kind of mode shadow), which is coincident with the power exchange mechanism that continuously transforms/converts one state into the other.

Finally, we study the case $k_x=\pi$ where, in general, we observe no transport of the gaussian beam [see Figs.~\ref{f3}(d1) and (d2)] due to the zero velocity at that region. However, for $\Delta\beta\approx 6$ we observe a splitting of both beams due to the excitation of the Dirac cone phenomenology with divergent plane waves propagating to the left and to the right across the lattice. The center of mass of both beams is shifted to the left for $\Delta\beta\lesssim6$ and to the right for $\Delta\beta\gtrsim6$, as shown in Fig.~\ref{f3}(d3). It is interesting to observe how this simple model could show an enhanced transport at $k_x=\pi$, something that is completely absent in standard 1D lattices~\cite{rep1,rep2}.

\section{Edge states and topology}

Up to now, we have looked for linear solutions of model (\ref{mo2}) by considering an infinite lattice, which naturally misses edge or boundary states. If present, those edge or boundary states might be connected to the bulk properties (bulk-boundary correspondence) through a topological invariant~\cite{asboth16,ortmann15,hasan10}, or just be localized modes without topological origin as the Tamm states~\cite{tamm}. 

We proceed to numerically compute the linear spectrum by diagonalizing a lattice of $N=30$ sites, considering two modes per site, obtaining the linear spectrum presented in Fig.~\ref{f4}(a). We also include the information of the size of eigenmodes by coloring the eigenvalues depending on the participation number of corresponding eigenvectors. The participation number is defined as $R= (P_u+P_v)^2/\sum_n (|u_n|^4+|v_n|^4)$ and is large (lighter color) when eigenmodes are of the order of the system size and small (darker color) when eigenmodes are of the order of a few lattice sites. We notice the appearance of two localized linear edge states at the middle of the gap for $\Delta\beta=0$, which are shown in Figs.~\ref{f4}(b1) and (b2). The state located at the left border has a flat phase (unstaggered) structure with equal sign for both modes, while the edge state located at the right border has opposite signs. Both states are very well localized close to the edge with $R=1.8$. For an increasing value of $\Delta\beta$, we observe that the edge mode become broader with $R=4.5$ for $\Delta\beta=2$. The profiles at this region are shown in Figs.~\ref{f4}(c1) and (c2), where we observe that edge states have a staggered phase structure with equal sign at the left border, while a staggered phase structure with opposite sign at the right edge. The same profile structure is observed for $\Delta\beta=4$ in Figs.~\ref{f4}(d1) and (d2), with $R=9.5$. After the gap closing transition (for $\Delta\beta\gtrsim 6$), we observe no edge states at all, and all linear modes become delocalized, belonging exclusively to the upper and lower bands.

The reported behavior of the edge states warrants further examination of the topological invariants, which provide a smoking gun for their topological nature.

\subsection{Topological invariants}

\begin{figure}[t]
\begin{center}
\includegraphics[width=0.48\textwidth]{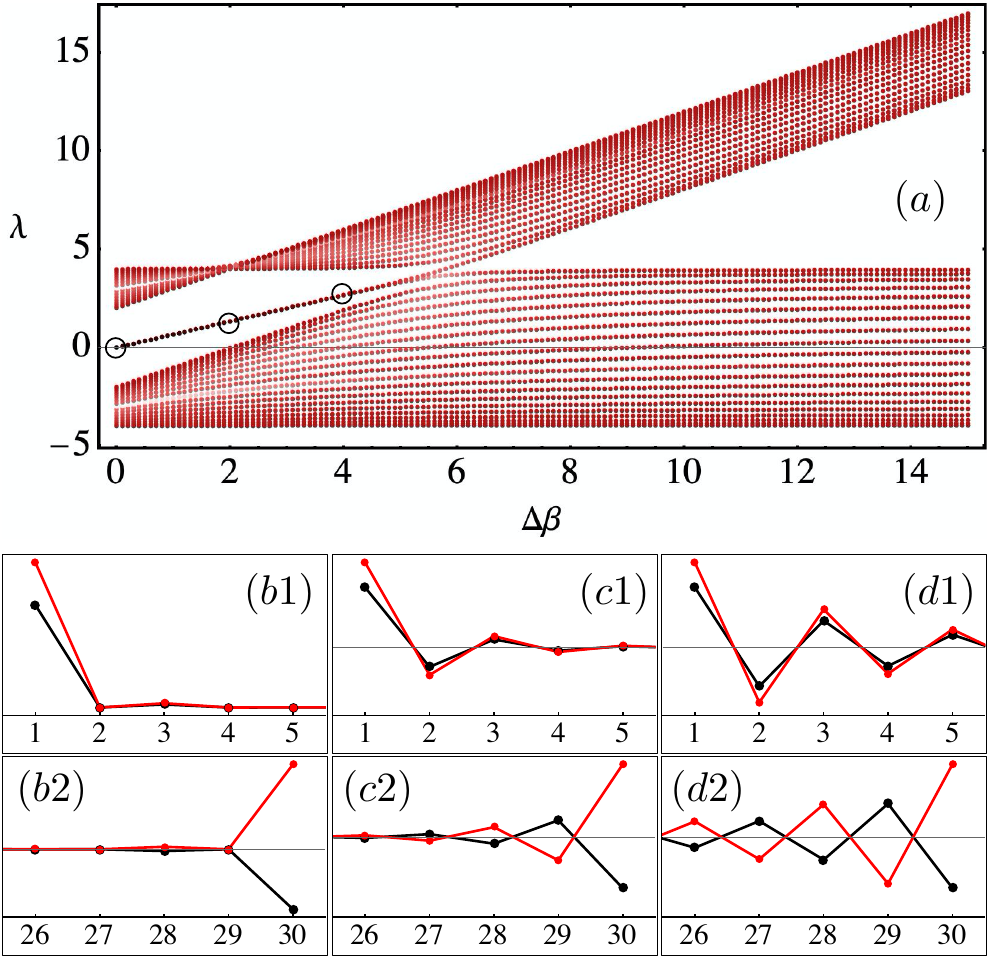}
\caption{(a) $\lambda$ versus $\Delta\beta$ for $N=30$. The participation ratio is plotted using different colors for each eigenvalue. (b1) [(b2)], (c1) [(c2)] and (d1) [(d2)] show the edge state profiles at the left (right) border for $\Delta\beta=0$, $2$ and $4$, respectively. Red and black lines correspond to the ``s'' and ``p'' mode amplitudes.$\{C_s,C_p,C_{sp}\}=\{1,-2,1.5\}$.}\label{f4}
\end{center}
\end{figure}

The edge states appear or disappear one each side of the gap closing point in Fig.~\ref{f4}(a). Here we proceed to compute the relevant topological invariant, the Zak phase~\cite{asboth16,atala13,ssh}, to unveil whether they are topological states or trivial localized states.

We will compute the Zak phase as a topological quantity that could help us to predict if our edge states have or not a topological origin. This phase is defined as $$\mathcal{Z}_{\pm} = i \oint \psi_{\pm}^* \cdot \frac{\partial\psi_{\pm}}{\partial k_x} dk_x\ ,$$ where $\psi_{\pm}$ represents the eigenmode for a given $k_x$-value for ``s'' and ``p'' modes, and for the upper ($+$) and lower ($-$) bands. We first define functions $g_1(k_x)\equiv \Delta\beta+2(C_s-C_p)\cos (k_x a)$ and $g_2(k_x)=2[C_2\exp(i k_x a)+C_1\exp(-i k_x a)]$, where $C_1\equiv C_{sp}$ and $C_2\equiv -C_{sp}$. After some algebra, we write in a compact form the eigenmodes of our lattice as $$\psi_{\pm}=\frac{1}{\sqrt{1+T_{\pm}^2}} \left(\begin{array}{c} 1 \\T_{\pm}\exp(i\phi)\end{array}\right)\ ,$$ with $T_{\pm}\equiv |g_2|/(g1\pm\sqrt{g_1^2+|g_2|^2})$, $\tan\phi=((1-\eta)/(1+\eta))\tan (k_x a)$, with $\eta\equiv C_1/C_2$. By grouping these expressions in the Zak phase formula, we obtain $$\mathcal{Z}_{\pm} = i \oint i\cdot z_{\pm}\cdot \frac{\partial \phi}{\partial k_x} dk_x\ ,$$ with $z_{\pm}\equiv T_{\pm}^2/(1+T_{\pm}^2)$. We compute the term related to the phase $\phi$ and take the limit $\eta\rightarrow -1$:

\begin{eqnarray}
\lim_{\eta\to -1}\frac{\partial \phi}{\partial k_x}=\lim_{\eta\to -1}\frac{1-\eta^2}{1+\eta^2+2\eta\cos(2k_x a)}=\nonumber\\ =\left\{\begin{array}{c l} \infty, &\ \ \mathrm{if}\ \ k_x a=\{0,\pi,2\pi,...\}. \\ 0, &\ \ \mathrm{otherwise.}\end{array}\right.
\end{eqnarray}

So, when integrating in the first Brillouin zone, this term will behave as a sum of two Dirac Delta functions: $-\pi [ \delta(k_x)+\delta(k_x-\pi/a) ]$. Now, as we have two bands in our lattice, we need to check the addition and subtraction of Zak phases. The addition must be zero modulo $2\pi$~\cite{asboth16}, so we first compute the sum obtaining directly that $z_+ + z_-=1$. Therefore, by evaluating $\partial \phi/\partial k_x$, in the first Brillouin zone, the addition of Zak phases simply gives 
\begin{equation}
\centering
\hspace{3cm}\mathcal{Z}_+ +\mathcal{Z}_-= \Delta \phi=2\pi\ ,\
\label{plus}
\end{equation}
which is constant as expected, as a good indication of the performed calculation. Now, we study the subtraction of phases. First of all, we notice that $z_+ - z_-=-g_1/\sqrt{g_1^2+|g_2|^2}$; therefore, we only need to evaluate this expression at $k_x=0,\pi/a$, obtaining that
\begin{eqnarray}
\mathcal{Z}_+ -\mathcal{Z}_-= -\pi \left.\frac{g_1}{|g_1|}\right|_{k_x=\{0,\pi/a\}} =\nonumber\\ =\left\{\begin{array}{c l}
0,& \mathrm{if}\ \ 0<\Delta\beta<2(C_s-C_p) . \\
-2\pi,& \mathrm{if}\ \ \Delta\beta>2(C_s-C_p).
\end{array}\right.
\label{minus}
\end{eqnarray}
Now, we are ready to find the Zak phases for each band in order to detect if there is any non trivial topological transition. By simply summing and subtracting (\ref{plus}) and (\ref{minus}), we finally obtain
\begin{eqnarray}
\mathcal{Z}_+= \left\{\begin{array}{c l}
\pi,& \mathrm{if}\ \ 0<\Delta\beta<2(C_s-C_p). \\
0,& \mathrm{if}\ \ \Delta\beta>2(C_s-C_p).
\end{array}\right.\nonumber\\
\mathcal{Z}_-= \left\{\begin{array}{c l}
\pi,& \mathrm{if}\ \ 0<\Delta\beta<2(C_s-C_p). \\
2\pi,& \mathrm{if}\ \ \Delta\beta>2(C_s-C_p).
\end{array}\right.
\label{zak}
\end{eqnarray}
Therefore, we clearly observe a nontrivial transition of $\pi$ in the Zak phase at a critical value of $\Delta\beta=2(C_s-C_p)$, which corresponds to the gap closing and which marks the onset of edge states. Therefore, we find that the edge states of this lattice with orbital degrees of freedom are topological.


Let us deepen our discussion. A first point has to do with the Altland-Zirnbauer classification of non-interacting topological phases~\cite{schnyder08}. In one-dimension the existence of topological bands require imposing symmetries on the system, as a one-dimensional system without symmetry constraints cannot have topologically non-trivial bands. Typically, this is reflected as topological bands protected by, for example, chiral or inversion symmetry as in the Su-Schrieffer-Heeger (SSH) model~\cite{asboth16}. Since model (\ref{mo2}) breaks chiral symmetry, one might wonder which symmetry is at play here. To clarify this, let us rewrite the Hamiltonian in terms of Pauli matrices:
${\cal H}={\vec d}({\vec k})\cdot {\vec \sigma} + d_0 {\cal I}$, where ${\vec d}({\vec k})=(0,2ig_2(k_x),g_1(k_x)/2)$. We see that although the z-component of ${\vec d}$ does not vanish, the x component does. This enforces the required symmetry. The topological invariant can be seen as the winding number of ${\vec d}$ in the $y-z$ plane. Therefore, the results obtained in this way are consistent with those presented above.

Finally, we note that one can also use a unitary transformation to make $d_z=0$, this is just a $\pi/2$ rotation around the ${\vec y}$-axis. The resulting Hamiltonian has equal $k$-dependent on-site energies and hoppings, which now include intra-cell and nearest neighbors hoppings.

\section{Flat band limit}

\begin{figure}[t]
\begin{center}
\includegraphics[width=0.48\textwidth]{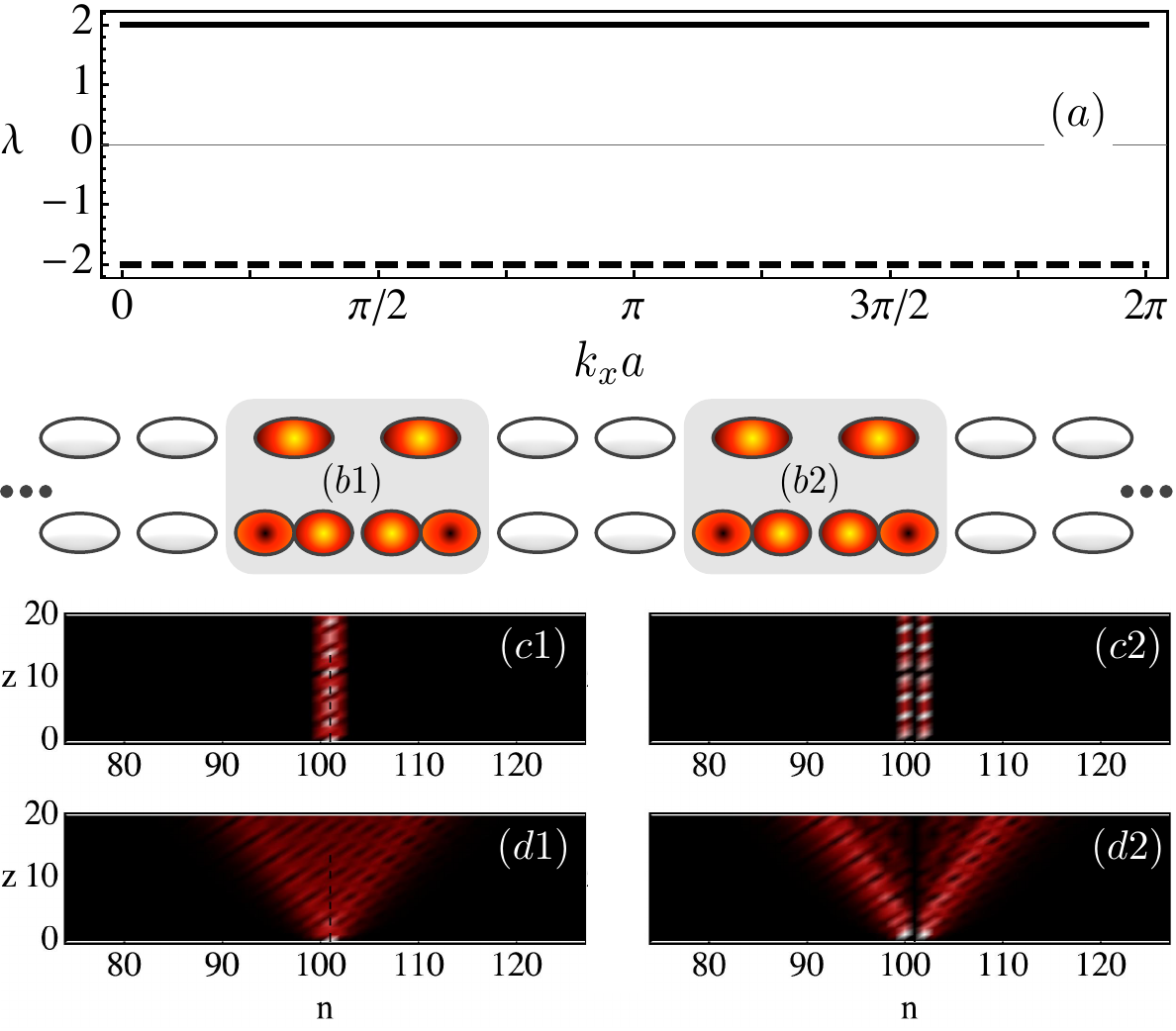}
\caption{(a) $\lambda$ versus $k_x a$ for $\Delta\beta=0$. (b1) and (b2) show the FB modes for $\lambda=2 C_{s}$ and $-2C_s$, respectively. (c1) and (c2) show the propagation of ``s'' and ``p'' modes, respectively, for an input condition $u_n(0)=A\delta_{n,no}$, $v_n(0)=0$, for $\Delta\beta=0$. (d1) and (d2) Same than (c) but for $\Delta\beta=1$. $C_s=-C_p=C_{sp}=1$.}\label{f5}
\end{center}
\end{figure}

By inspecting the linear spectrum (\ref{dr1}), we realize that there is a way to cancel the dependence on $k_x$ for both bands, what means getting a flat band spectrum. This condition requires that $\Delta\beta=0$, $C_p=-C_s$ and that $C_{sp}^2=-C_s C_p = C_s^2$. Therefore, by assuming a perfectly symmetric coupling condition $C_s=-C_p=C_{sp}$, we are able to reduce the linear bands to two completely flat ones located at $\lambda=\pm 2 C_{s}$. The states related to these bands have a profile exactly equal to the eigenmodes of a dimer system, with profiles $\{u_n,u_{n+1},v_n,v_{n+1}\}=\{A,A,-A,A\}$ for $\lambda=2 C_{s}$ and $\{-A,A,A,A\}$ for $\lambda=-2C_s$ (with $A$ any given amplitude). We show the spectrum at the FB limit in Fig.~\ref{f5}(a), including the FB modes in Figs.~\ref{f5}(b1) and (b2).
%
%
These states can be excited everywhere on the array and occupy only two lattice sites. These perfectly spatially localized states correspond to FB linear compactons, having an exactly zero background. It is important to notice that although a $\Delta\beta=0$ condition is not experimentally possible on a strict 1D lattice [Fig.~\ref{f1}(a)], this limit could be implemented in the ribbon configuration [Fig.~\ref{f1}(c)] by correctly tuning the propagation coordinates.

An interesting property for this lattice is that both initially dispersive bands become completely flat with no dispersive modes neither transport across the system (this is similar to recent reported phenomena on Aharonov-Bohm caging~\cite{ABC1,ABC2}). Therefore, by tuning the coupling or detuning coefficients it would be possible to effectively transform the system from a conductor to an insulator. In Fig.~\ref{f5}(c1) and (c2) we show the propagation of a single-site fundamental input excitation for a detuning coefficient $\Delta\beta=0$. First of all, we observe how the intermode coupling excites the dipolar mode. Then, both modes propagate well localized in space, although the input condition was not an exact FB mode profile (when injecting any of the two FB states, they simply remain localized along the $z$ direction). If we, for example, vary the detuning coefficient to $\Delta\beta=1$ [see Fig.~\ref{f5}(d1) and (d2)], a completely different dynamics is observed due to the acquired curvature of bands. This transition is quite similar to the one experience in Sawtooth lattices, although in that system only one band becomes completely flat for a critical ratio between coupling constants~\cite{OLsaw}.

\section{Conclusions}

Here we have studied a minimal case of a one-dimensional linear photonic lattice containing besides the fundamental state, one more orbital mode. We showed how the coupling between these orbital states leads to a wealth of phenomena including a transition from a regime where the bands become flatter exhibiting reduced transport to one where they become more dispersive. By tuning the detuning between orbital states we have also shown the possibility of inducing a completely flat linear spectrum with compact localized states. Furthermore, our characterization of the states in the bands in terms of the Zak phase shows a topological transition at a gap closing point in parameter space accompanied by the ensuing topological edge states. Due to its simplicity, our proposed lattice could become an interesting candidate for realistic topological photonic operations in all-optical technologies. We hope that our results could further motivate the use of orbital degrees of freedom in photonic lattices. We anticipate further experimental work in this direction.

\section*{Acknowledgements}
The authors acknowledge Dr. Magnus Johansson for useful discussions at the beginning of this work. This work was supported in part by Programa ICM Millennium Institute for Research in Optics (MIRO) and FONDECYT Grants No.1191205 and No.1170917. LEFFT acknowledges support from The Abdus Salam International Centre for Theoretical Physics and the Simons Foundation.


\begin{thebibliography}{99}
\bibitem{com1} Depending on the excitation wavelength, refractive index contrast and waveguide dimensions, a waveguide could operate selectively in single- or multi-mode regimes~\cite{dipole1}.
\end{thebibliography}
\end{document}